\documentclass[pre, onecolumn ]{revtex4}

\usepackage{graphicx}
\usepackage{epstopdf}
\usepackage{amsmath}
\usepackage{amssymb}
\usepackage{mathrsfs}
\usepackage{euscript}     
\usepackage{bm}
\usepackage{verbatim}   
\def\be{\begin{equation}}
\def\ee{\end{equation}}
\def\beu{\begin{equation*}}
\def\eeu{\end{equation*}}

\def\bsub{\begin{subequations}}
\def\esub{\end{subequations}}
\def\ie{i.e.}

\providecommand{\stext}[1]{\text{\tiny{#1}}}  

\providecommand{\inv}{^{-1}}

\def\smallhalf{\tfrac{1}{2}}

\providecommand{\mean}[1]{\left\langle#1\right\rangle}

\begin{document}
\noindent

\title{A comment on `Wave-breaking Limits for relativistic electrostatic waves in a one-dimensional warm plasma' [\textit{Phys. Plasmas} \textbf{13} 123102]\\by R. M. G. M. Trines and P. A. Norreys
}
\author{A. E. Charman}
\email{acharman@physics.berkeley.edu}
\author{R. R. Lindberg}
\affiliation{Department of Physics, University of California, Berkeley, Berkeley, CA 94720, USA}
\date{\today}
\maketitle

In \cite{trines:2006}, the authors state that a principle aim was ``to create some order in the diverse landscape of warm fluid models and establish which model is most suitable for the study of wave breaking in thermal plasma.''  Unfortunately, we must conclude that they have obfuscated more than they have clarified, and have advanced the waterbag model only by exaggerating or misstating its benefits while mischaracterizing features of alternative warm fluid theories.

The cold 1D wave-breaking limit is singular in both the colloquial and mathematical senses of the word, because a number of physical and mathematical characterizations all change simultaneously and abruptly at a single, well-defined value for the electric field amplitude.  In particular, a plasma wave amplitude corresponding to the cold nonlinear wave-breaking amplitude\cite{akhiezer:1956, akhiezer:1956b}, given by
$E_{\stext{WB}} = E_0 \sqrt{2(\gamma_{\phi} - 1)}$,
where $E_0 = \tfrac{m c \omega_p}{e}$ is the cold linear wave-breaking field, $c$ is the speed of light \textit{in vacuo}, $\omega_p = \sqrt{\tfrac{4\pi n_0 e^2}{m}}$ is the linear plasma frequency in Gaussian units for electrons of mass $m$, charge $-e$, and density $n_0$, and $\gamma_{\phi} = \tfrac{1}{\sqrt{ 1- v_{\phi}^2}}$ is the relativistic kinematic factor corresponding to the plasma wave phase velocity, is simultaneously associated with:
onset of microscopic trajectory crossing;
breakdown of the single-valuedness of the fluid velocity;
appearance of multi-stream flow;
onset of particle trapping by the Langmuir wave;
appearance of a singularity (divergence) in the density and in the slope of the electrostatic field;
mathematical invalidation of the fluid theory;
breakdown of periodicity/appearance of additional Fourier content;
onset of the loss of wave coherence;
and achievement of the maximum sustainable electric field for a wave of well-defined phase velocity.

In contrast, \textit{any} non-zero amount of thermal motion is sufficient to ensure that these properties of cold wave-breaking will no longer occur simultaneously, if at all.  Some features may emerge gradually rather than suddenly, while others may arise at different wave amplitudes, and still others may no longer occur.

In the warm case, one must therefore adopt a single criterion to define wave-breaking.  In the most literal sense of the phrase, wave-breaking as typified by shallow water waves, in which multi-stream macroscopic flow becomes manifest, occurs only in the cold 1D limit of plasma waves.  In a warm fluid theory, crossing of the individual electron trajectories does \textit{not} necessarily invalidate a fluid description, or lead to a breakdown of the single-valuedness of the fluid velocity, as long as the local ensemble of particle velocities remains consistent with the moments of the distribution function presumed by the fluid theory.

Any attempt to define a sharp wave-breaking limit in terms of particle trapping or loss of periodicity is problematic, because in the warm, relativistic case, such effects will, strictly speaking, appear at any non-zero value of the plasma wave amplitude, and will grow gradually and continuously as the wave amplitude grows, with no sharp threshold or critical value apparent.  Continual trapping in the tails of a thermal distribution is pointed out by Trines and Norreys (TN) themselves, which makes their definition of wave-breaking all the more puzzling.

Warm wave-breaking is believed to be associated both with a balance between space charge forces and pressure effects, and also with actual particle-trapping, but any specific connections between the wave-breaking limit and trapping must be investigated within a kinetic description which accurately models the latter, and will be revealed \textit{a posteriori}.   Attributing wave-breaking \textit{a priori} to some arbitrarily chosen trapping fraction, or to the trapping of particles with some specified cutoff in \textit{initial} velocity, as is done by TN, is essentially arbitrary and can result in an \textit{ad hoc} definition for wave-breaking with little physical meaning or interest.  Trapping and related loading effects, where trapped particles can deplete and/or de-phase the collective plasma oscillation, are of course of interest, but this does not imply that we can \textit{a priori} precisely equate the achievable upper bound on plasma wave amplitude with any specific value for any particular quantitative measure of these trapping or loading effects.

A wave excited to (or just below) wave-breaking may or may not appear to lose coherence, but attempts to rigorously define the coherence of the plasma wave are also problematic.  For any wave, the degree of coherence is a statistical measure, conventionally defined in terms of the mean and covariances of the wave field, but the electric field which appears in the Vlasov description is already the mean field, not the local microscopic field.  For plasmas in or very near thermal equilibrium, it is possible to bootstrap information about the \textit{fluctuations} in the field from a Vlasov model, by using detailed balance while regarding each electron in turn as a test particle moving in a background Vlasov plasma consisting of the remaining electrons, but it is not clear whether these notions can be extended to the case where a large-amplitude collective Langmuir oscillation is present, and the electron plasma is not even in local thermodynamic equilibrium (LTE).

Although we can offer no airtight mathematical proof that shocks or other singularities do not eventually arise from some initial conditions in either Vlasov or fluid theories at any non-zero temperature, no experimental or theoretical evidence suggests that any shocks appear or singularities arise, at least before the maximum achievable field at a given phase velocity is reached.  At any non-zero temperature, no mathematical or physical quantities necessarily diverge at wave-breaking, so divergences cannot be used reliably to define warm wave-breaking.

Conventionally, wave-breaking has been instead defined simply as the maximum attainable amplitude for a traveling plasma wave of specified phase velocity.  This is the definition used by Akheizer and Polovin\cite{akhiezer:1956} for cold relativistic plasma waves; it turns out to be equivalent to definition used by Dawson \cite{dawson:1959} for wave-breaking in cold, non-relativistic plasmas; and it is identical to the definition used by Coffey\cite{coffey:1971} for non-relativistic waves in a warm plasma, and that used by by Katsouleas and Mori\cite{katsouleas:1988} (KM) for ultra-high phase-velocity, relativistic plasma waves in a warm plasma.  It was therefore also the definition adopted by Schroeder, Esarey, and Shadwick (SES)\cite{schroeder:2005} for their treatment of wave-breaking of plasma waves of arbitrary phase velocity in moderately warm plasmas.  This definition of the wave-breaking limit in terms of the maximum achievable field  has a long-standing claim of historical priority, has a precise operational meaning, is of direct experimental interest in the context of plasma-based acceleration schemes, leads to a unique, well-defined answer within most dynamical models, and seems to agree well with fully kinetic simulations.  Given these facts, this definition deserves to be retained, and any non-equivalent definitions or characterization should be given some other name.

Instead of following the long-standing definition, TN attempt to quantify a ``kinetic'' definition of wave-breaking in terms of the trapping of those electrons whose \textit{initial} speed is given by the 3D RMS thermal speed $v_{\stext{th}} = \sqrt{\tfrac{3 T_e}{m}}$, where here $T_e$ is the electron temperature in energy units.  But $v_{\stext{th}}$ is just a representative RMS value, so this definition appears to provide illusory precision without true accuracy.  Why should one take the thermal velocity as the precise cutoff, and not half or twice this value?  In particular, why should one consider the 3D speed limit $v_{\stext{th}}$ as an appropriate cutoff rather than the RMS thermal velocity in the direction of the 1D wave, namely $\tfrac{v_{\stext{th}}}{\sqrt{3}}$?

TN offer an incorrect  justification for equating wave-breaking with the trapping of particles exactly at  $v_{\stext{th}}$.  They claim that, because according to the Clemmow-Wilson generalization of the Bohm-Gross dispersion relation, relativistic or non-relativistic plasma waves at reasonable temperatures always have a linear phase speed in excess of $v_{\stext{th}}$,  it somehow follows that electrons with \textit{initial speed} $v \le v_{\stext{th}}$ will contribute to the collective plasma wave oscillation, while electrons with speed $v > v_{\stext{th}}$ will not.  As a matter of logic, this syllogism is flawed because it would be equally true for, say, $\smallhalf v_{\stext{th}}$, or $10^{-2} v_{\stext{th}}$, or any other chosen cutoff speed less than or equal to the thermal speed.  As a matter of physics, this argument is also flawed.  As explicitly demonstrated in the pioneering work of Bohm and Pines\cite{pines:1952}, the division between single-particle and collective degrees-of-freedom is meaningful in a plasma for waves or other structures with spatial length-scales larger than about one Debye length $\lambda_e = \sqrt{\tfrac{T_e}{4\pi n_0 e^2}},$ but even then this distinction is not sharp, and it cannot in general be extended to a corresponding division in velocity space, based on the thermal speed as cutoff between collective and single-particle effects.  If instead the thermal velocity $v_{\stext{th}}$ is replaced in this argument with the phase velocity $v_{\phi}$, can it even be true that particles with \textit{initial} velocity $v < v_{\phi}$ contribute to the electric field of the wave, but particles with \textit{initial} velocity $v > v_{\phi}$  do not?  No, because this would essentially lead to a prediction that effects of particles which can \textit{eventually} be trapped by the plasma wave can already be neglected \textit{initially}, independent of the amplitude of the wave and whether they are actually trapped or not.  What if instead we consider instantaneous velocities rather than initial velocities?   It is true that particles with an instantaneous velocity equal to or greater than the Langmuir wave \textit{phase velocity} $v_{\phi}$  tend to be trapped in the plasma wave potential and can deplete the wave by what can be interpreted as nonlinear Landau damping or inverse Cerenkov radiation, but even in that case it  cannot be true in general that the trapped particles ``do not contribute'' to the plasma wave.  From the analysis of Bernstein-Greene-Kruskal (BGK) nonlinear modes\cite{krall:1973}, we know in one spatial dimension that for an essentially arbitrary choice of distribution function for the untrapped electrons, the distribution of trapped electrons can be chosen to reproduce almost any periodic Vlasov potential $\Phi(x).$  This implies that in general even trapped particles can contribute significantly, in fact crucially, to the collective space-charge fields.  In the context of plasma-based acceleration, obviously trapped particles contribute (albeit deleteriously) to the collective field via beam-loading.  Simply put, the source for the Vlasov  potential is always proportional to the distribution function integrated over all momenta, not just those less than some prescribed cutoff.  Generally speaking, theories which remove trapped particles or totally disregard their contribution to the collective electric field upon becoming trapped can actually perform worse than theories which do not incorporate trapping effects in the first place.

Despite these unconvincing arguments in support of the ``kinetic'' definition of wave-breaking, it seems that it  is adopted by TN primarily in order to avoid a finite limit on the wave-breaking amplitude in the ultra-relativistic limit $v_{\phi} \to c$, and to justify, in a circular fashion, the use of the waterbag model.  In regards to plasma-based accelerator applications, the authors state that
``it would be problematic if there would exist a finite upper bound for the wave-breaking limit in a warm plasma even for $v_{\phi} \to c$'', without, however, specifying exactly how or to whom this would pose a problem.  We fail to see why or how this poses a problem, either mathematically, conceptually, or experimentally.  Langmuir waves with truly luminal phase velocities cannot be excited by either laser beams of sub-luminal group velocity or particle drive beams of finite energy.   In fact, TN emphasize the so-called ultra-relativistic regime, for which $\gamma_{\phi}^2 \tfrac{T_e}{m c^2} \gg 1$, but for typical laser-plasma parameters, the group velocity of the laser and hence the phase velocity of the plasma wave will be relativistic in the sense that $1 - v_{\phi} \ll 1,$ but definitely not ultra-relativistic, since $\gamma_{\phi}^2 \tfrac{T_e}{m c^2} < 1$.

Even if a wave of $v_{\phi} \to  c$ could somehow be achieved through some other excitation mechanism,  it would still be very surprising if a plasma of finite density and non-zero but finite temperature could sustain coherent electric fields of arbitrarily large amplitude, and neither current accelerator experiments nor envisioned accelerator applications rely on any such possibility.  Even in the cold relativistic case, luminal phase velocities could only be achieved by plasma waves containing infinite energy density.  For a finite number of electrons in some finite interaction volume, each with finite energy, we fail to see how anything but a finite \textit{macroscopic} mean field can be achieved, or why this is a shortcoming.  For a given drive beam in a plasma of given density and temperature, a certain relativistic but sub-luminal phase-velocity wave of a certain amplitude can be excited, and, in any plasma-based accelerator scheme, energy gain beyond that which can be provided by such a plasma wave would require a different configuration of drive and plasma, or additional acceleration stages.  This is not an insurmountable problem, but a widely acknowledged fact.  The maximum achievable plasma wave amplitude is of course of great interest, but it is not necessarily ``problematic''  in any application merely because it turns out to be finite at any phase velocity, as long as it is adequately large.

In order to find a divergent wave-breaking limit as $v_{\phi} \to c,$ TN proceed to adopt an unphysical choice for the momentum distribution function, namely a waterbag model with sharp 1D momentum boundaries initially at the 3D thermal momenta, \ie, $\pm p_{0}$, where $p_0 \approx m  v_{\stext{th}}$ is the momentum associated with the 3D thermal velocity, assuming the latter is non-relativistic.  Through a bit of mathematical accident (the same factor of $\tfrac{1}{\sqrt{3}}$ arising in different places for completely different reasons), the RMS \textit{longitudinal} speed of this distribution then matches that of a 1D Maxwellian.  Given TN's definition for ``kinetic wave-breaking,''  the waterbag has been chosen in  circular fashion, as the simplest possible distribution for which no trapping whatsoever occurs until particles of initial velocity $v = v_{\stext{th}}$ are trapped, which has been asserted by fiat as marking the onset of wave-breaking.  It is not surprising that they then conclude this is in a sense the best distribution by which to study wave-breaking according to their definition.  But conversely, if their definition of wave-breaking is rejected as \textit{ad hoc}, there remains little reason to resort to a waterbag model except possibly mathematical simplicity.

To better understand these issues and address the claims by TN against alternative fluid models,
we should first pause to clarify the distinction between kinetic and fluid descriptions, in the context of underdense plasmas.  In a true plasma, kinetic energies dominate over inter-particle potential energies, collisions may be neglected over relevant time-scales, and the uncorrelated (mean field) Vlasov kinetic description involving the reduced single-particle phase space distribution function is generally adequate.  In contrast, a fluid description involves the evolution of a reduced set of functionals of the Vlasov distribution function, usually given by certain phase space moments of the conditional momentum distribution at each point in space.  This distinction between fluid and kinetic theories is not perfectly sharp, but is still useful.  With more moments included, the full distribution function can be reproduced with more accuracy.  Likewise, if many different ``species'' of fluids corresponding to different slices of momentum space are evolved, then the fluid description becomes essentially kinetic.  Conversely, in practice, and especially numerically, one can only evolve a finite amount of information about the Vlasov distribution function even in what is regarded as a kinetic description.

Effectively, if at each point in space, the conditional momentum distribution function is parameterized by a small number (in absolute terms) of moments or other phase space quantities, then then we say the description is fluid-like, while it is kinetic if the momentum space parameterization is more finely-grained.  Also, the division is not always mutually exclusive.  For example, any solution of the cold fluid equations is also an \textit{exact} solution of the Vlasov equation, since by assumption the momentum distribution starts off infinitesimally narrow, and without collisions it will remain so.

Strictly speaking, we should also distinguish between hydrodynamic and non-hydrodynamic fluid theories.  Ideal plasmas generally satisfy the Bogoliubov hypothesis postulating a separation of scales between the range of interaction (Debye length), mean free path between collisions, and system size,
so at (but only at) sufficiently long times (compared to the inverse  collisional frequency $\nu_e\inv$), 
it is assumed that collisional processes establish a LTE, and the hydrodynamical variables are sufficient to describe the system and parameterize the distribution function.  For most laser-plasma scenarios, this time-scale is much longer than the plasma period as well as relevant total interaction times, and the plasma remains essentially collisionless.  In these so-called initial or kinetic stages, \textit{hydrodynamic} observables can be employed in part or whole to describe approximately the state of the collisionless plasma, but the evolution is not hydrodynamic, because collisions are absent.  Fluid theories in this regime must rely on other physics for their validity and their closure.

TN admit that the  waterbag model is ``mostly hydrodynamic in nature,'' yet hint that it is somehow capturing certain kinetic features.  We argue that it is \textit{entirely} fluid in nature where it is well-defined.  As long as the the upper and lower boundaries of the momentum distribution remain single-valued, \ie, before any trapping, the waterbag remains an exact solution of the Vlasov equation, as follows from Liouville's Theorem.  But any such solution is completely characterized by a central momentum and a momentum spread at each spatial point, which is a quintessential example of a fluid theory.  In the presence of any actual trapping, the assumption of single-valuedness of the momentum space boundaries must break down.  The waterbag model cannot describe trapping, and in fact invalidates its own assumptions and breaks down mathematically at the onset of any trapping, whereas  this very sharp threshold for trapping would be unphysical because of the unrealistic sharp boundaries in momentum space.

While advocating the waterbag model, TN completely mischaracterize the covariant warm relativistic fluid theory of Schroeder, Esarey and Shadwick (SES) \cite{schroeder:2005}, and by extension the related but distinct Hamiltonian fluid theory developed by Shadwick, Tarkenton, and Esarey (STE) \cite{shadwick:2004,shadwick:2005}.  At one point, TN claim that such warm fluid models are based on the assumption that ``the plasma temperature is much smaller than the kinetic energy associated with the mean flow,'' suggesting the relevant small expansion parameter is something like $\tfrac{T_e}{m c^2 \left[\mean{\gamma}-1\right]}$.  This is obviously incorrect, as it would not even hold for a plasma in thermodynamic equilibrium with no bulk flow.  In SES and STE, it is clearly stated that the relevant small parameter is essentially $\tfrac{T_e}{m c^2}$, the ratio of the temperature (thermal energy) in the average rest frame to the rest mass energy, which is indeed very small for realistic laser-plasma parameters.
 
Then TN assert that SES themselves claim their theory is a ``representation of an adiabatic process'' rather than a ``low-density/low temperature expansion.''   Never do SES or STE claim their theories are anything but leading-order \textit{asymptotic} expansions, essentially in the relative rest-frame temperature parameter $\tfrac{T_e}{m c^2}$, about the cold fluid theory.  No expansion in density need be made directly, and all that is required is that for given temperature the density is sufficiently low so that pair correlations remain small and the Vlasov equation is valid.  No adiabatic equation of state is assumed or implied.  STE only point out that the in the case of quasi-static plasma response to a short intense laser pulse, ``qualitatively the results agree with thermodynamic arguments for an adiabatic process,'' in the sense that the axial momentum spread naturally increases where the plasma is compressed and decreases where it is rarified.  In the general case including the effects of laser quiver, no \textit{scalar} equation of state is presumed nor can be derived for these theories, for they allow for the anisotropy that naturally arises in the presence of a transverse laser field.

Also, TN accuse these authors of confusing the mathematical break-down of the SES fluid theory with the physical breaking of the wave.  But the warm fluid theories of SES or STE do \textit{not} break down mathematically at the maximum achievable field.  No shocks, divergences, or other singularities appear.  Nor is there any ``soft'' breakdown in the sense of an invalidation of the assumptions on which the fluid theories are based.  The local rest-frame temperature increases in regions of compression, but only by a factor $< O(10)$ for typical laser-plasma parameter regimes, so the asymptotic expansion and related moment closure remain valid even at wave-breaking.  Even in the unphysical limit $v_{\phi} \to c,$ the relative effective temperature stays below unity.
On the contrary, it is the waterbag fluid model of TN which breaks down mathematically as the wave-breaking amplitude is reached, because the assumption of single-valued momentum boundaries will be invalidated as electrons at one of those boundaries are trapped.  TN claim that ``great care must be taken in verifying that the singularity of any warm fluid model used for the hydrodynamic definition of wave-breaking is applied judiciously so that its physical meaning is retained,'' and insinuate that the model of SES somehow fails in this regard.  We are at a loss to understand what this means, since: unlike the waterbag model, the SES model exhibits no singularities at wave-breaking; and furthermore all physical observables are ultimately finite, so infinities or singularities cannot have physical meaning or ever be ``applied judiciously'' when they do appear.

Also, TN  claim that in subsequent work \cite{schroeder:2006}, the authors confuse single-particle orbits with fluid motion, and further that they erroneously use the potential from a cold plasma calculation to track particles in a warm plasma.  This is incorrect on both counts.  Nowhere do Schroeder, \textit{et al}. suggest that the particle-tracking is equivalent to the fluid calculation.  Instead they use particle tracking to independently assess and confirm the accuracy of their fluid-based model, and then to \textit{deduce} the \textit{approximate} level of particle-trapping associated with wave-breaking in a particular parameter regime.  Electrostatic fields from a cold plasma are used as a perfectly reasonable approximation in the single-particle tracking because, as is pointed out, the macroscopic fields from 1D cold fluid, warm fluid, and full Vlasov simulations approximately agree.  The resulting trajectories for test charges will therefore also approximately agree.

Finally, TN repeatedly assert that the fluid theory that is used by SES to determine the wave-breaking limit is invalid because it fails to satisfy the ``fundamental inequality'' of Taub.  This too is entirely incorrect.  This inequality\cite{taub:1948} follows as a direct consequence of the Cauchy-Schwarz inequality applied to moments of the stress-energy tensor.  It was intended to rule out certain \textit{ad hoc} relativistic equations of state that were invoked without being derived from moments.  But if appropriate moments are \textit{consistently} calculated or estimated from \textit{any} well-defined  distribution function (which is relativistically-invariant, non-negative, and normalizable), then Taub's inequality should be satisfied automatically, and in fact it is always satisfied as an equality in the relativistic  theory of SES provided one correctly identifies the ``energy density'' and ``pressure'' terms in the inequality with the associated physical moments, and calculates all asymptotic quantities consistently to the same order.  Again, if one starts with any consistent distribution function and calculates moments from it, rather than invoking \textit{ad hoc} relations between moments, then Taub's inequality will be vacuously satisfied, and tells us nothing new.

It is straightforward to explicitly verify that this inequality is satisfied by SES.  Immediately following his equation (4.4), Taub writes this inequality in the form:
\be\label{eqn:taub1}
  \tfrac{(\rho^0)^2}{m^4 c^4} w \left(-T_\alpha^{\phantom{\alpha}\alpha} \right) \ge \tfrac{(\rho^0)^4}{m^4}.
\ee
In the notation and with the ordering assumptions of SES (\ie, \textit{consistently} neglecting third-order terms), we have the following correspondences:
$\left(\rho_0\right)^2 \rightarrow m^2 n_p^2$,  $w \equiv m^2 T_{\alpha\beta}\tfrac{U^\alpha U^\beta}{(\rho^0)^2} \rightarrow 
  	m^3 c^2 T_{\mu\nu}\tfrac{J^\mu J^\nu}{m^2 n_p^2} = mc^2\tfrac{n_p^2}{h},$ and
 $-T_\alpha^{\phantom{\alpha}\alpha} \rightarrow mc^2 T_\mu^{\phantom{\mu}\mu} = mc^2 h.$
 It can then be shown that both sides of the Taub inequality \eqref{eqn:taub1} are equal, and equal to $ c^2\left(\frac{n_p}{h} - 1 \right).$  TN appear to erroneously conclude otherwise either by misidentifying the terms or making inconsistent Taylor expansions.  The (3D) pressure term $p$ of Taub is related to quantities defined in SES  as  $p \equiv \tfrac{1}{3}\left( w + T_\alpha^{\phantom{\alpha}\alpha}\right) \rightarrow  \tfrac{1}{3}m c^2 \left( \tfrac{n_p^2}{h} - h \right)$, while Taub's energy per unit mass $\epsilon$ is identified with $\epsilon \equiv \frac{1}{\rho^0} w - c^2 \rightarrow  \tfrac{1}{m n_p} mc^2\tfrac{n_p^2}{h} - c^2 = c^2\left( \tfrac{n_p}{h} - 1 \right).$
 
Not only is the waterbag model really a fluid theory rather than a kinetic theory, it is not generally expected to be an especially physically realistic or accurate one.  In the theory of SES, retention of first-order and second-order moments and neglect of third-order moments can be shown to be essentially equivalent to assuming a Gaussian for the conditional momentum distribution at each spatial location, which is the distribution of \textit{maximum entropy} consistent with these given averages and covariances.  Compared to the Gaussian, for the same RMS momentum spread the waterbag distribution has an entropy lower by $\smallhalf + \smallhalf \ln\left[2\pi \right] - \ln\left[2\sqrt{3}\right]$ natural units (nats), or approximately $0.2546$ bits, per electron.   This may not sound like much, until we remember that the electrons are assumed uncorrelated, so the overall entropies and entropy differences are extensive.  Consider a typical Ti:sapphire laser-plasma configuration with a laser of wavelength $1 \, \mu\text{m}$ and focused spot size of about $40\, \mu\text{m}$ and a $3\, \text{mm}$ interaction length inside a plasma of density $n_0 \sim 10^{18} \, \text{cm}^{-3}.$  Then $O(10^{11})$ electrons may participate in the plasma wave.  From a Bayesian, or information-theoretic interpretation the entropy, this means that if we really only possess reliable information up to the second-order moments, but use the waterbag instead of the Gaussian, we are pretending to about $O(10^{10})$ bits of information that in practice we do not in fact possess, equivalent to a medium-sized library of about $O(10^4)$ volumes.  From a more conventional frequency or combinatoric perspective, this means that for every way an initial frequency distribution with given first and second moments near the waterbag could be generated by the proverbial monkeys throwing electrons into phase space cells, a distribution near the Gaussian could be generated in more than $10^{10^9}$ ways.
 
The Vlasov-Poisson model constitutes a deterministic dynamical system which is first-order in time, so obviously any solution is only as good as the initial conditions.  Simply put, the waterbag model starts with inaccurate and unrealistic initial conditions of unjustifiably low entropy.  It is not surprising that such an unphysical distribution function might subsequently do unphysical things.

In fact, it is only through the work of STE that we understand why, despite its unphysical assumptions, the waterbag model can work at all, yielding answers that are close to those of the SES theory in reasonable parameter regimes: asymptotically, the leading order thermal corrections to the cold fluid model only depend on the momentum distribution function through the first and second-order moments, and therefore \textit{any} distribution function with the same first-order moments and centered second-order rest-frame moments will lead to the same leading-order thermal corrections.

TN claim that the quasi-static waterbag model is ``implicitly favored over other distributions'' which can only ``approximately'' satisfy two conditions that are ``exactly'' satisfied by the waterbag, namely that ``the effect of small amounts of trapped particles on the unbroken wave is neglected'' together with the condition that ``propagation is an adiabatic process.''  Again we are puzzled by what this might mean.  No single-species fluid theories, whether ultimately based on Gaussian, waterbag, or other unimodal distribution functions, really describe trapped particles at all.  Conversely, even within a kinetic (Vlasov) distribution, \textit{any} truncated distribution will defer particle-trapping in the same way the waterbag distribution does.  Also, fluid theories like SES or STE, consistently derived from the collisionless Vlasov equation essentially by assuming a Gaussian momentum distribution, have no viscous heating effects, and possess a local conservation law for Boltzmann entropy.  Entropy density can be advected by the flow from one place in space to another, but entropy cannot be created during compression or rarefaction, and thermal energy will not flow spontaneously from a warmer to a colder region in any way that increases entropy (at least not without the introduction of some additional course-graining not present in the Vlasov equation).

Despite the counter-arguments of TN, considerations of entropy do reveal that the waterbag model is indeed  less ``general''  than the Gaussian distribution, in the precise and objective sense that it presumes far more initial information about the particle momenta than in typical scenarios one will in fact possess.   Of course, as $\gamma^2 \beta_{\stext{th}}$ becomes larger, we expect that any fluid theory that tracks only a small number of low-order momentum moments may suffer in accuracy.  If this really is the regime of interest (which seems highly doubtful at least for laser-plasma or plasma-based acceleration applications), the answer is not to continue to assume either a waterbag or a Gaussian model with specific functional relations between higher and lower-order moments, but rather to evolve independently some additional number of these higher-order moments.  Closure will no longer emerge in a distribution-independent manner, but in principle, such a scheme can be closed by considerations of maximum entropy, equivalent to assuming the momentum distribution function remains in the exponential family to which the Gaussian belongs, although in practice, this may lead to computational inefficiencies, so other approximations may be needed.

The only remaining grounds on which to favor the waterbag model over the SES or STE models might be mathematical convenience, tractability, or simplicity.  While the equations of motion for particle density, momentum, and momentum spread might be somewhat easier to derive in the waterbag model as compared to the SES theory, the resulting equations are not particularly simpler to solve.  In fact, TN do not even calculate an exact expression for the ``kinetic'' wave-breaking amplitude as defined by them, but only upper and lower bounds on it.  They claim that these upper and lower estimates ``only differ'' by an amount proportional to $\tfrac{\ln{2}}{2} \left\{\tfrac{m c^2}{T_e}\right\}^{1/4}$ in the present notation, where $T_e$ is the initial temperature, but  once again $\tfrac{T_e}{m c^2} \ll 1$ in typical regimes, so this difference is not especially small.  Actually, the exact form for the limit within the waterbag model can be determined analytically in terms of elliptic functions, but is not particularly simpler or easier to work with than the exact calculations for the wave-breaking field within the warm fluid theory in the quasi-static case.

Compared to the waterbag model, the warm-fluid theory of SES  is at least as simple mathematically, does not make unphysical presuppositions about additional low-entropy structure in the distribution function, and does not exhibit any singularities or divergences at or below the conventional fluid definition for wave-breaking of a plasma wave of given phase velocity in a warm plasma.  Contrary to the claims of TN, it does not presume \textit{a priori} any particular adiabatic equation of state, its moments begin and remain consistent with the Taub inequality, and for typical parameters appropriate to laser-plasma applications, evolution also remains consistent with its own fundamental assumptions about the smallness of the momentum spread in the local rest frame.


\end{document}